\RequirePackage[2020-02-02]{latexrelease}
\documentclass[prl,showpacs,preprintnumbers,amsmath,amssymb]{revtex4}

\usepackage[dvips]{graphics}
\usepackage{tikz}

\begin{document}

\title{Quantum mechanics with real numbers: entanglement, superselection rules and gauges}

\author{Vlatko Vedral}
\affiliation{Clarendon Laboratory, University of Oxford, Parks Road, Oxford OX1 3PU, United Kingdom}

\date{\today}

\begin{abstract}
\noindent We show how imaginary numbers in quantum physics can be eliminated by enlarging the Hilbert Space followed by an imposition of - what effectively amounts to - a superselection rule. We illustrate this procedure with a qubit and apply it to the Mach-Zehnder interferometer.
The procedure is somewhat reminiscent of the constrained quantization of the electromagnetic field, where, in order to manifestly comply with relativity, one enlargers the Hilbert Space by quantizing the longitudinal and scalar modes, only to subsequently introduce a constraint to make sure that they are actually not directly observable. 
\end{abstract}


\maketitle                           

\section{Introduction}

We have all seen statements of the type (I am not quoting from any particular text here, just rephrasing things that have appeared in articles and books): ``complex numbers are fundamental in quantum physics. The Schr\"odinger equation is a diffusion equation, but in complex time, and the ``i" in ``$i\hbar \partial/\partial t$" is unavoidable. Yes, we do use complex numbers to represent the phase of any classical wave, $e^{i(kx-\omega t)}$, but this is just a matter of convenience in our calculations. Ultimately, at the end, we always have to take the real part of this expression to arrive at a measurable entity. In quantum physics, on the other hand, it is impossible not to use complex numbers." 

Dyson puts it dramatically: ``...But then came the surprise. Schr\"odinger put the square root of minus one into the equation, and suddenly it made sense. Suddenly it became a wave equation instead of a heat conduction equation. 
…And that square root of minus one means that nature works with complex numbers and not with real numbers." \cite{Dyson}.There are even recent papers claiming experimental evidence that rules out quantum physics with real numbers only, \cite{Renou,Chen}.

In this paper, I would like to present a way of eliminating the need for complex numbers in quantum physics and show how to only ever use the reals. The idea is old \cite{Stuckelberg,Myrheim}, though my exposition is, I believe, novel. 

The price to pay for avoiding the imaginary components will be, first and foremost, the need to dimensionally enlarge the representation of the states as well as of the observables (the trick that is overlooked by the above mentioned experiments claiming complex numbers to be indispensable). In addition, we will need to restrict the number of observables in this higher representation, so that we achieve the full equivalence with quantum mechanics. Without the latter step, we would obtain a theory that is larger than quantum physics and therefore could also be ruled out as we will see in what follows.

Neither of these procedures, the enlargement of the space, and the subsequent restrictions, are, of course, foreign to quantum physics. The canonical quantization of the electromagnetic field in the Lorenz gauge is the best-known example \cite{Scharf}. In the remaining part of the paper we will proceed as follows. We first present a general method for eliminating the imaginary numbers from quantum physics. We then apply this to the Mach-Zehnder interferometer. The next step is to present a single qubit as two qubits in the real representation in order to exhibit the role played by entanglement. A comparison is then made with the Gupta-Bleuler method of quantizing the electromagnetic field \cite{Gupta}. Finally, we discuss the implications of our results and speculate on possible extensions of quantum physics.

\section{The Main Construction}

We will show this with a two-level system, and the generalisation to any dimensionality is straightforward. First of all, it is clear that the general superposition $a|0\rangle + b|1\rangle$, can always be encoded by $4$ real numbers written as a vector $(a_r,a_i,b_r,b_i)$, where the subscripts indicate the real and imaginary parts of $a$ and $b$. 

The key observation now is the following one and it is the only additional trick one needs to understand. If the amplitude of either of the states acquires a phase factor equal to $\sqrt{-1}$ (i.e. a $\pi/2$ phase shift) then this amounts to the transformation $a_r\rightarrow a_i, a_i\rightarrow -a_r$ and likewise for $b$. Therefore, multiplication by the imaginary number $i$ simply can be represented by the matrix whose action is as follows:
\begin{equation}
\begin{bmatrix} 
	0 & -1 \\
	1 & 0 \\
	\end{bmatrix}
\begin{bmatrix} 
	a_r \\
	a_i \\
	\end{bmatrix}=
\begin{bmatrix} 
	-a_i \\
	a_r \\
	\end{bmatrix}
\end{equation} 
The most general dynamics is given by the Schr\"odinger equation and it is easy to see how to now represent it with real numbers only. The prescription is that all imaginaries are substituted by the above matrix (which is equal to $i\sigma_y$) and all unit numbers by the two-by-two identity matrix.  
The Schr\"odinger equation for a qubit is in the Pauli basis given by
\begin{equation}
i\hbar \frac{d}{dt} 
\begin{bmatrix} 
	a \\
	b \\
	\end{bmatrix}= \left(h_0
\begin{bmatrix} 
	1 & 0 \\
	0 & 1 \\
	\end{bmatrix}+h_1
\begin{bmatrix} 
	0 & 1 \\
	1 & 0 \\
	\end{bmatrix}+h_2
\begin{bmatrix} 
	0 & -i \\
	i & 0 \\
	\end{bmatrix}+h_3
\begin{bmatrix} 
	1 & 0 \\
	0 & -1 \\
	\end{bmatrix}\right)
\begin{bmatrix} 
	a \\
	b \\
	\end{bmatrix}
\end{equation}
which, following the above prescription (after putting the $i$ on the right hand side of the equation and absorbing it into the $4$ matrices), becomes in the real representation:
\begin{equation}
\hbar \frac{d}{dt} 
\begin{bmatrix} 
	a_r \\
	a_i \\
   b_r \\
	b_i \\
	\end{bmatrix}= \left(h_0
\begin{bmatrix} 
	0 & 1 & 0 & 0 \\
   -1 & 0 & 0 & 0 \\
   0 & 0 & 0 & 1 \\
	0 & 0 & -1 & 0\\
	\end{bmatrix}+h_1
\begin{bmatrix} 
	0 & 0 & 0 & -1 \\
   0 & 0 & 1 & 0 \\
   0 & -1 & 0 & 0 \\
	1 & 0 &  0 & 0\\
	\end{bmatrix}+h_2
\begin{bmatrix} 
	0 & 0 & -1 & 0 \\
   0 & 0 & 0 & -1 \\
   1 & 0 & 0 & 0 \\
	0 & 1 &  0 & 0\\
	\end{bmatrix}+h_3
\begin{bmatrix} 
	0 & 1 & 0 & 0 \\
   -1 & 0 & 0 & 0 \\
   0 & 0 & 0 & -1 \\
	0 & 0 & 1 & 0\\
	\end{bmatrix}\right)
\begin{bmatrix} 
	a_r \\
	a_i \\
   b_r \\
	b_i \\
	\end{bmatrix}
\end{equation}
By construction, this equation leads to the same dynamics as the original Schr\"odinger equation. For instance, we can obtain the usual Larmor precession by setting $h_0=h_1=h_2=0$ and $h_3=\hbar \Omega$. In this case we obtain the following equations $\dot a_r = a_i, \dot a_i = - a_r, \dot b_r = -b_i, \dot b_i = b_r$. If the initial state is such that $a_r(0)=b_r(0)=1/\sqrt{2}$ and $a_i(0)=b_i(0)=0$, the solution is $a_r (t) =\cos \Omega t/\sqrt{2}, a_i (t) =\sin \Omega t/\sqrt{2}, b_r (t) =\cos \Omega t/\sqrt{2}, a_i (t) =-\sin \Omega t/\sqrt{2}$. In the standard qubit representation this is simply $(|0\rangle + |1\rangle)/\sqrt{2} \rightarrow (e^{i\Omega t}|0\rangle +e^{-i\Omega t} |1\rangle)/\sqrt{2}$, i.e. the phase between the two states precesses at the rate $2\Omega$ as expected.

The generalisation to higher dimensional systems is immediate, since the same prescription applies in that case too. This also includes the infinite dimensional systems such as the quantum simple harmonic oscillator. One should also be aware of the fact that the imaginary component could have higher dimensional representations than $2$, although we will not be using it in this work. We will come back to the infinite dimensional systems when we discuss the quantization involving gauges. 

Now, however, a comment is in order regarding the fact that the real representation contains much more than the original quantum mechanics. Namely, if we look at all the possible linear transformations of the vector $(a_r,a_i,b_r,b_i)$, then one such transformation is what is known as the universal NOT gate. This transformation sends the quantum state $a|0\rangle + b|1\rangle$ into its orthogonal  state $b^*|0\rangle - a^*|1\rangle$, which in the real representation leads to the state
$(-b_i,b_r,-a_i,a_r)$. The matrix that achieves this in the real representation is $\sigma_z\otimes i\sigma_y$. In ordinary quantum physics, the universal not is an anti-linear operation and not representable by a unitary transformation. There are, of course, many other operations that are not allowed in quantum mechanics, even the innocently looking operation that conjugates both of the amplitudes (the basis for the partial transpose criterion for entanglement, which itself is not a physical operation). An easy way of eliminating all such operations is to insist that all allowed operations must commute with $i\sigma_y \otimes  I$ (the universal NOT fails to commute because the Pauli $y$ and $z$ matrices do not commute).

In fact, the full basis is given by the following matrices: $ i\sigma_y\otimes I,i\sigma_y\otimes \sigma_x,I\otimes -i\sigma_y, i\sigma_y\otimes \sigma_z$. This is the real representation of the identity and the Pauli matrices. The reason is that in ordinary quantum theory, the identity commutes with all other operations, however, in real quantum mechanics the identity becomes $(-i)I$ which is represented by $i\sigma_y\otimes I$. The rationale behind this will become even clearer when we use the entangled two qubit representation of a single qubit. 

This brings us to a surprising statement whose validity has been known for a long time \cite{Stuckelberg,Myrheim}: a commuting subspace of the full real theory, gives us back quantum physics in the same way that a commuting subspace (say with $\sigma_z$) of quantum physics leads us to classical physics (i.e. the physics where superpositions of the state $|0\rangle$ and $|1\rangle$ are not allowed). One therefore wonders if the real extension of quantum theory could be thought of as a possible gateway to going beyond.

\section{The Mach-Zehnder Qubit}

Here is a good place to introduce another, frequently seen, argument for the necessity of using complex numbers in quantum mechanics. One can think of any two-input-two-output interferometer as a computation performing a NOT gate on the input state, or something similar (probably a Mach-Zehnder interferometer is better described as performing a $-I$ operation, namely introducing a $\pi$ phase shift in both states of the interferometer). 

Now, quantum dynamics is continuous, which means that it is meaningful to talk about the transformation taking us half way through the interferometer. This is then a square root of NOT or of $-I$ (since applying the transformation twice should lead us to the full interferometric transformation). Both of these transformations clearly require imaginary numbers if we remain within the $2$ by $2$ matrix representation. 

However, we saw that in the $4$ dimensional representation, this is no longer the case. We now treat the evolution of the Mach-Zehnder action using this representation. The simplest is to first write the three transformations (the first beamsplitter, the two mirrors and the last beamsplitter) using the conventional Pauli basis:
\begin{equation}
\begin{bmatrix} 
	1 & i \\
	i & 1 \\
	\end{bmatrix}
\begin{bmatrix} 
	0 & i \\
	i & 0 \\
	\end{bmatrix}
\begin{bmatrix} 
	1 & i \\
	i & 1 \\
	\end{bmatrix}
\end{equation}
which multiplies out to the negative identity as indicated before. We now follow the enlargement prescription so that $1\rightarrow I$ and $i\rightarrow -i \sigma_y$. The Mach-Zehnder transformation is then
\begin{equation}
\begin{bmatrix} 
	1 & 0 & 0 & -1 \\
   0 & 1 & 1 & 0 \\
   0 & -1 & 1 & 0 \\
	1 & 0 &  0 & 1\\
	\end{bmatrix}
\begin{bmatrix} 
	0 & 0 & 0 & -1 \\
   0 & 0 & 1 & 0 \\
   0 & -1 & 0 & 0 \\
	1 & 0 &  0 & 0\\
	\end{bmatrix}
\begin{bmatrix} 
	1 & 0 & 0 & -1 \\
   0 & 1 & 1 & 0 \\
   0 & -1 & 1 & 0 \\
	1 & 0 &  0 & 1\\
	\end{bmatrix}
\begin{bmatrix} 
	a_r \\
   a_i \\
   b_r \\
	b_i \\
	\end{bmatrix}=
\begin{bmatrix} 
	-a_r \\
   -a_i \\
   -b_r \\
	-b_i \\
	\end{bmatrix}
\end{equation}
and, clearly, all transformations contain real numbers only. 

The above calculation involves one effective qubit, but we must be careful when generalising real quantum physics to two or more qubits \cite{Myrheim}. The problem exists even when each qubit on their own obeys quantum physics. For instance, the action of applying a phase to the first or the second of two qubits is indistinguishable in ordinary quantum physics. The state $i|\psi\rangle \otimes |\phi\rangle$ gives the same results in any experiment as the state $|\psi\rangle \otimes i|\phi\rangle$. However, in real quantum mechanics this corresponds to two different states that could in principle be discriminated (for the simple reason that the state is given by $4$ real numbers). To make things worse, they are also different from the states $|\psi\rangle \otimes |\phi\rangle$ and $i|\psi\rangle \otimes i |\phi\rangle$.

This presents a problem for the real quantum mechanics and it signals that we need to reduce the space of states even further. The solution is similar to how we deal with multi-mode fermionic states. When we have two fermions, we can only do full operations in spaces with different parity. Namely, there is a superselection rule that prohibits us from superposing the vacuum state with the one fermion state. Therefore, we have to confine ourselves either to the subspace where we have the vacuum and two fermions (one in each mode) or where we have the vacuum in one mode and a fermion in the other one. These two subspaces are of different parity and any unitary operation is allowed in each, but the two can never be connected physically. In other words, each subspace is equivalent to a qubit, but together we still only have one qubit and not two.  

\section{Entanglement} 

The real representation used here could be thought of as a ``second quantization" of the amplitudes pertaining to quantum states. Namely, the real vector $(a_r,a_i,b_r,b_i)$ can be written in the form of a tensor product of two qubits as
\begin{equation}
\left(
\begin{array}{c}
a_r\\
a_i\\
\end{array}
\right) \otimes
\left(
\begin{array}{c}
1\\
0\\
\end{array}
\right) +
\left(
\begin{array}{c}
b_r\\
b_i\\
\end{array}
\right) \otimes
\left(
\begin{array}{c}
0\\
1\\
\end{array}
\right) 
\end{equation}
From this representation, it becomes clear why the multiplication by $i$ is encoded as $i\sigma_y\otimes I$ for it is only the first of the two qubits that is changed (the real and imaginary parts of $a$ and $b$ get swapped and the real part acquires a minus sign). 

Furthermore, this state can be interpreted as an entangled state of two qubits, where we note that the states of the first qubit have not been normalised. When normalization is included, the state, written in the Dirac notation, is given by:
\begin{equation}
|a|\left(\frac{a_r}{|a|} |0\rangle + \frac{a_i}{|a|} |1\rangle\right)|0\rangle+|b|\left(\frac{b_r}{|b|} |0\rangle + \frac{b_i}{|b|} |1\rangle\right)|1\rangle \; .
\end{equation}
As we saw, if all quantum operations were allowed on this state, then the resulting single qubit dynamics would contain more possibilities than contained in quantum physics. 
In fact, this two qubit representation of a single qubit allows us to see exactly why imposing the restriction to operations that commute with $i\sigma_y\otimes I$ is needed
to recover the ordinary quantum physics. It is because the imaginary $i$ has been upgraded to a two-by-two matrix in the real representation, but in ordinary quantum mechanics it is still just a number that ought to commute with everything else. 

Let us, for the time being, ignore this fact and proceed to compute the amount of entanglement in this state by treating it like any other two-qubit quantum state. The simplest way is to trace out the second qubit and
take the entropy of the first qubit reduced state. This state is
\begin{equation}
\rho_1=|a|^2 \left(\frac{a_r}{|a|} |0\rangle + \frac{a_i}{|a|} |1\rangle\right)\left(\frac{a_r}{|a|} \langle 0| + \frac{a_i}{|a|} \langle 1|\right) + |b|^2\left(\frac{b_r}{|b|} |0\rangle + \frac{b_i}{|b|} |1\rangle\right)\left(\frac{b_r}{|b|} \langle 0| + \frac{b_i}{|b|} \langle 1|\right)
\end{equation}
which, when written in the maxtix form yields
\begin{equation}
\rho_1=
\begin{bmatrix} 
	a^2_r + b^2_r & a_ra_i +b_rb_i \\
	a_ra_i +b_rb_i & a^2_i + b^2_i  \\
	\end{bmatrix}
\end{equation}
The eigenvalues are $r_1 = 1/2(1+\sqrt{1-4\det(\rho_1)})$ and $r_2=1-r_1$. The entropy of entanglement is therefore equal to $E=-r_1 \ln r_1 -r_2 \ln r_2$.

Maximum entanglement occurs when $a_r =b_i$ and $a_i = -b_r$. For instance, the qubit state $e^{\pi/4}|0\rangle + e^{-\pi/4}|1\rangle$ corresponds to the maximally entangled state in the real two qubit representation (where the state is $(|0\rangle + |1\rangle)/\sqrt{2}\otimes |0\rangle + (|0\rangle - |1\rangle)/\sqrt{2}\otimes |1\rangle$). Disentangled states (product states since we are confined to the globally pure ones) on the other hand are those where either $a=0$ or $b=0$, or when $a_r=b_r$ and $a_i=b_i$. 

Of course, due to the constraints imposed in recovering the usual quantum theory, this entanglement is only ``representational". This is because to confirm entanglement we need to be able to measure at least two different complementary observables on each subsystem. However, the constraint imposed in order to recover quantum mechanics, tells us that $\sigma_y$ is the only observable accessible on one of the qubits (the first one in our notation). One qubit is therefore effectively ``classicalised" in order to reduce the real theory to quantum theory. It is here that we see similarities with fermions, where two fermionic modes are equivalent to one qubit only. 

\section{Electromagnetic field quantization in the Lorenz gauge}

Here we would like to outline some parallels between the extension of quantum physics we have been discussing with the constrained quantization employed in the presence of gauges. If one wishes to be manifestly relativistically covariant, one needs to work in the Lorentz gauge, where all the $4$ modes of the vector potential are quantized. The quantum $A$-field is, in the Fourier expansion, given by
\begin{eqnarray}
A_\mu (x) & = & \int \frac{d^3 k}{\sqrt{2\omega_k (2\pi)^3}}\sum_{\lambda =0}^3 \Bigg(a_\lambda (k)\epsilon_\mu (k,\lambda) e^{-ikx} 
+  a^{\dagger}_\lambda (k)\epsilon_\mu (k,\lambda) e^{+ikx}\Bigg) \; ,
\end{eqnarray}
where $a$ and $a^\dagger$ are the usual annihilation and creation operators for the $4$ modes. We now have the subsidiary condition which guarantees the gauge independence (in the case of the free electromagnetic field):
\begin{equation}
\sum_\mu \frac{\partial A^+_\mu}{\partial x_\mu} |\Psi\rangle =0 \; ,
\end{equation}
where $A^+_\mu$ is the positive frequency part of the vector potential (the one containing the annihilation field operators). This condition, known as the Gupta-Bleuer constraint \cite{Gupta}, must hold at all space points and at all times. It ensures that the expectation value of $\partial A_\mu/\partial x_\mu$,  $\langle \Psi| \partial A_\mu/\partial x_\mu |\Psi\rangle$, vanishes at all points and times. In the momentum space this yields a simple constraint:
\begin{equation}
\sum_\mu k_\mu a_\mu |\Psi\rangle =0 \; ,
\end{equation}
in other words, any non-transverse components of the field are annihilated at all times. It turns out that for the free field (without charges whose treatment is not relevant here - see \cite{Heitler}) it is sufficient to postulate this just at the initial time. 

If, without loss of generality, we assume that $k_\mu = (\omega,0,0,\omega)$, then this condition reduces to 
\[
(a_3-a_0)|\Psi\rangle = 0 \; ,
\]
which states that physical states must have an equal number of longitudinal and scalar photons. This ensures that the longitudinal and scalar photons never contribute anything observable to any physical analysis, which is why they are frequently called ``ghosts". The average of any observable simply vanishes as far as the contribution from the $0$ (temporal) and $3$ (longitudinal $x$) modes. In other words, the probability of ever observing the longitudinal and scalar photons is identically zero throughout. Indeed, if the state $|\Psi^{\prime}\rangle$ is obtained from $|\Psi\rangle$ by an emission of a ghost-photon (and therefore by a gauge change whose potentials are proportional to $\partial_\mu \Omega$) it must have the form 
\[
|\Psi^{\prime}\rangle = (1 + \lambda \sum_\mu k_\mu a^{\dagger}_\mu)|\Psi\rangle \; .
\]
Now the overlap of this new state with ghost-photons with another physical one, $|\Psi_1\rangle$ is 
\[
\langle \Psi^{\prime}|\Psi_1\rangle = \langle \Psi|(1 + \lambda \sum_\mu k_\mu a_\mu)|\Psi_1\rangle = \langle \Psi |\Psi_1\rangle \; ,
\]
i.e., it is unaffected by the presence of the ghost-photons. It is, in fact, clear that any number of applications of the operator $\lambda^{\prime} \sum_\mu k^{\prime}_\mu a^{\dagger}_\mu$ must lead to the same physical state. A gauge change is now, at the operator level, given by $a_\mu\rightarrow a_\mu+\Omega (k)k_\mu$. This satisfies the gauge condition, because $\Omega$ satisfies the Lorenz condition $\sum_\mu \frac{\partial^2 \Omega (x)}{\partial x^2_\mu}  =0$, which in the momentum representation ($\Omega (x) = \Omega (k) e^{ikx}$) simply states that $k^2_\mu=0$, i.e. that photons are massless. 

The biggest drawback of the manifestly covariant quantisation of the electromagnetic field is that it leads to states with a negative norm. This is because the Lorentz invariance implies that the commutation relations between the $4$-vector $A$ potential components must have the Minkowski signature. Therefore, for one of the four modes - the scalar mode - the commutation relation between the annihilation and creation operators has a wrong sign, $[a_0,a^\dagger_0]=-1$. This implies that, if the vacuum state of that mode is properly normalised so that $\langle 0|0\rangle = 1$, then the first excited state (and all the odd states) have a negative norm because $\langle 1|1\rangle=\langle 0|a_0a^\dagger_0|0\rangle = -1 +\langle 0|a^\dagger_0a_0|0\rangle = -1$. 

A knock-on effect is also that the observables that are normally considered to be Hermitian, in the case of the scalar mode become anti-Hermitian. This is interesting since in our real representation of quantum physics, the Pauli operator $\sigma_y$ is multiplied by the $i$ in order to eliminate the imaginary numbers in the representation. The operator $i\sigma_y$ is anti-Hermitian which means that if one of the eigenstates of $\sigma_y$, with the eigenvalue of $-i$, is assumed to be positive norm, then the other eigenstate must have a negative norm. This is for the same reason as in the scalar mode since we can write $\sigma_y =-i\sigma_++i\sigma_-$.

To make the analogy with real quantum mechanics more transparent we introduce a single qubit but with an indefinite metric. This simply means that $\langle 0|0\rangle = 1= - \langle 1|1\rangle$. The norm of a general superposition $c|0\rangle +s|1\rangle$ is given by $c^2 - s^2$. If this is to be normalised, $c=\cosh x$ and $s=\sinh x$ (instead of the $\cos$ and $\sin$ as in the definite norm conventional quantum physics). The orthogonal state is given by $s|0\rangle +c|1\rangle$. 

It is clearly impossible to interpret the squares of amplitudes as probabilities, since they can exceed unity. The negative norm is the main problem, so how can one deal with it while preserving the indefiniteness of the metric? One way is to introduce another such indefinite qubit. Now it seems that the problem has doubled, however, we can choose to work in the subspace $|0\rangle\otimes|0\rangle, |1\rangle\otimes|1\rangle$ where the metric is always positive ($-1\times -1 = 1$). One can simply declare that the physical subspace is composed of the states $|\psi\rangle$ which obey the following constraint:
\begin{equation}
(\sigma_+\otimes \sigma_- + \sigma_-\otimes \sigma_+)|\psi\rangle = 0
\end{equation}
This resembles the fermionic superselection rule where one works in the positive parity subspace of two fermionic modes as explained above.  Also, more appropriately for the present discussion, it resembles the Gupta-Bleuler quantization in the Lorenz gauge where the constraint is arranged so that the scalar and the longitudinal modes excitations cancel out, $(a_3 - a_0)|\psi\rangle = 0$. We will say more on this point in the discussion below.

The constraint on the indefinite qubits, as in the case of the real quantum physics extension, boils down to the fact that two qubits with indefinite metric are equivalent to one qubit with a definite metric. 

We have now completed the formal part of the exposition (though there is much more to be explored here, see e.g. \cite{Wolff,Adler,Gull}) and would like to proceed with the discussion regarding possible physical implications of the ideas presented.

\section{Discussion}

Extending the space of states and operations to real Hilbert Spaces has the appeal that no imaginary number are required to describe either quantum states or quantum operations. After all, complex numbers are just ordered pairs of real numbers, so it might not be surprising to see that only the real numbers suffice.

This teaches us one important lesson, however. It is not the need for the complex numbers that makes quantum physics unusual; it is, instead, the need for the non-commuting elements of reality, or, what Dirac called, the $q$-numbers. Indeed, the way we got rid of the $i$ was to substitute it with an operator that need not commute with other relevant operators. This is why this trick to make quantum physics real, in fact, makes it even more ``quantum". Eliminating complex numbers is therefore not a return to the classical physics of real numbers, but quite the opposite. Quantum physics does not need complex numbers, but it does need $q$-numbers and the question as always is which classical entities to upgrade to $q$-numbers.

That extended theory, however, leads to impossible operations, such as the universal NOT. The enlarged theory needs to be restricted to recover the ordinary quantum theory. However, there is still a question regarding whether some of the extended operations could still be realised out there in Nature.

The intuition here comes from my recent work with Marletto \cite{Marletto,Marletto2}, in which we argued that the ghost modes of the electromagnetic field are both real (in the sense that they affect charges) and detectable (through coupling to charges). The idea is that the Coulomb forces are in the Lorenz gauge mediated by the scalar potential. It works in the following way. Every charge perturbs the scalar modes and creates coherent states in them. Formally, the state of the charge and the modes is an entangled state if a charge is in a superpositions of different locations (since the coherent states are centred around those locations). 

Given what we said about the Gupta-Bleuler constraint, it would seem that the charge-scalar-modes entanglement is only a formal kind of entanglement, of the same kind as the entanglement between two qubits in the real quantum mechanics. Namely, it would seem impossible to confirm this entanglement since the photons in the scalar mode are undetectable. This, however, is not true. The argument that Marletto and I have presented involves another charge and the reduction in the coherence of the first charge is a direct witness of entanglement. 

Is something like this possible for real extensions of quantum physics? More precisely, could two qubits be entangled through the extension into the real domain? In order to mimic as closely as possible the story about the electromagnetic gauges, we imagine a real qubit in the entangled representation as two qubits, and then coupled to another real qubit of the same type. However, we will allow these two real qubits to couple to one another only through the $i\sigma_y\otimes i\sigma_y$ interaction between the ``amplitude" qubits. The question is if the two real qubits can now ever become entangled if they initially start in a product state. The answer superficially seems to be a yes, because the same interaction could entangle two normal qubits. However, in the real quantum mechanical representation, the matrix $i\sigma_y$ implements the multiplication by $i$. In that sense, our interaction only produces a global phase and has no entangling power. Of course, if the Hamiltonian also contained $\sigma_z$ terms, say signifying the energy of the extension qubits, then this would, in combination with $i\sigma_y\otimes i\sigma_y$, be able to generate various entanglements. Something like that happens in the scalar modes, where both $a^\dagger a$ and $a+a^\dagger$ - which do not commute - play a role. 

Thinking along these lines, one direction for testing for effects going beyond quantum physics would be to probe if things ``fail to commute even more" than stipulated by the standard quantum theory. If, as we have explored, the quantum amplitudes also end up being $q$-numbers, then the effects would automatically be testable (e.g. the standard phases would become non-Abelian as in the case of anyons \cite{Nayak} or quaternionic amplitudes \cite{Peres}).     

In conclusion, attempts to phrase quantum physics on real Hilbert Spaces can, in fact, lead to theories broader than quantum physics. A constraint then has to be used to recover (the ordinary) quantum physics, however, the ``leakage" into the larger Hilbert Space can always - at least in principle - be detected \cite{Peres}. It seems to the author that a search for such deviations is certainly a worthwhile enterprise. 

\textit{Acknowledgments}: The author is grateful to the Moore Foundation and the Templeton Foundation for supporting his research.

\end{document}